\documentclass[twocolumn,showpacs,preprintnumbers,amsmath,amssymb]{revtex4}
\usepackage[dvips]{graphicx} % para incluir gráficos
\begin{document}
\preprint{IMAFF-RCA-05-06}
\title{Quantum theory of an accelerating universe}

\author{Pedro F. Gonz\'{a}lez-D\'{\i}az and Salvador Robles-P\'{e}rez}
\affiliation{Colina de los Chopos, Centro de F\'{\i}sica ``Miguel
Catal\'{a}n'', Instituto de Matem\'{a}ticas y F\'{\i}sica
Fundamental,\\ Consejo Superior de Investigaciones Cient\'{\i}ficas,
Serrano 121, 28006 Madrid (SPAIN).}
\date{\today}
\begin{abstract}
We review some of the well-known features of quantum cosmology,
such as the factor ordering problem, the wave function and the
density matrix, for a dark energy dominated universe, where
analytical solutions can be obtained. For the particular case of
the phantom universe, we suggest a quantum system in which the
usual notion of locality (non-locality) of quantum information
theory have to be extended. In that case, we deal also with a
quantum description where the existence of a non-chronal region
around the big rip singularity is explicitly accounted for.
\end{abstract}


\maketitle

\section{Introduction}

Besides confirming that the universe expands in an accelerated
fashion, recent data coming from SNIa and other observations
\cite{Riess98}\cite{Spergel03} leave the issue of the precise way
in which such an accelerated behavior actually occurs unsettled.
In fact, the possibility for a superaccelerated expansion beyond
what is predicted by a cosmological constant has been raised,
implying serious theoretical difficulties. Several models have
been proposed so far in order to explain accelerated expansion.
The most popular among them are the so-called quintessence models,
which are characterized by a universe filled with an homogeneous
fluid with an equation of state $p=w \rho$, where $p$ and $\rho$
are the pressure and the energy density of the fluid,
respectively, and $w$ is a constant within the range
$-1<w<-\frac{1}{3}$, being the value of $w=-1$ equivalent to a
cosmological constant. This would trigger by itself an exponential
expansion if one assumes no other energy or matter sources in the
universe. This acceleration would correspond to a value for $w$
less than $-1$, so allowing for what is called the phantom regime.
This regime entails violation of the dominant energy condition,
and might imply some interesting features from the point of view
of the quantum theory. Some of such features will be analyzed in
this paper within the realm of a more general formalism where
other possible scenarios are also contemplated.

On the other hand, from a quantum mechanical standpoint, the
universe is a rather special system since it cannot be described
as a whole in terms of space-time coordinates but in terms of
geometries. Hence, it offers a particularly interesting framework
to deal with some well-known issues of quantum mechanics, such as
those related with the notion of non-locality.

In this paper we shall consider therefore the quantum theory of an
accelerating universe which is filled with dark energy, both when
the dominant energy condition is satisfied and for vacuum contents
where such a condition is manifestly violated. In the latter case
the notion of non-locality is discussed in a multiverse scenario,
where it must be necessarily generalized or extended. The
generalized quantum theory of Hartle \cite{Hartle90} is then
applied to cases where the future singularities are replaced for
non-chronal bounded regions.

The paper can be outlined as follows. Sec. II deals with a phantom
universe covering the entire time interval in such a way that it
becomes describable as a multiverse. In sec. III we review the
canonical Hamiltonian formalism for a quantum universe,
particularizing in the problems related with the density matrix
and the possibility for the existence of entangled states in the
phantom multiverse. The generalized quantum theory is applied to
the quantum multiverse in which the big rip singularity is
replaced for a bounded non-chronal region in sec. IV, where a
decoherence function is used and observable probabilities are
isolated from it using the Hartle procedure. We summarize and
conclude in sec. V. An appendix on the orthogonality properties of
the wave function is also added.

\section{The phantom multiverse}

\begin{figure}[h]
\begin{center}
\includegraphics[width=9cm]{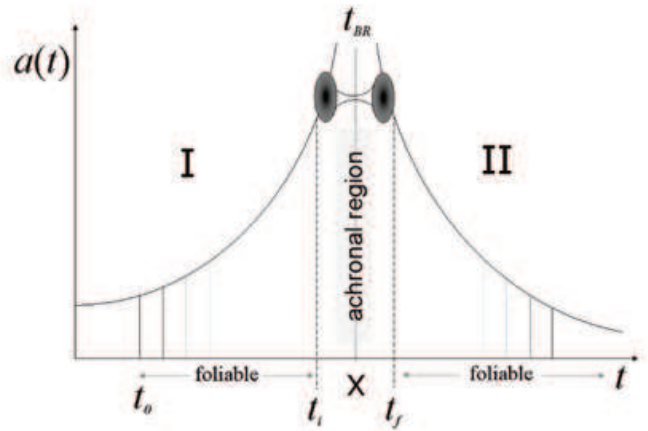}
\end{center}
\caption{Wormholes could connect both sides of the universe
avoiding that way the big rip singularity. For times $t$, $t_{i} <
t < t_{f}$, i.e., the achronal region, the manifold cannot be
foliated.} \label{univ}
\end{figure}

For a flat Friedmann-Lema\^itre-Robertson-Walker (FLRW) universe
filled only with dark energy, the equations of motion can be
obtained from the Hamiltonian constraint in the phase space,
\begin{equation}\label{Hamiltonian Constraint}
\mathcal{H} = - \frac{2 \pi G}{3} \frac{p_a^2}{a} + \rho_0 a^{-3
w} = 0,
\end{equation}
where $p_a$ the canonical momentum,  $G$ is the gravitational
constant, $\rho_0$ is the energy density at the coincidence time,
i.e., the time in which the dark energy started to dominate the
expansion of the universe, and $a$ is measured in units of $a_0$.
In deriving Eq. (\ref{Hamiltonian Constraint}) we have used the
integrated form of the energy conservation law,
\begin{equation}
d\rho = -3 (p+\rho) \frac{da}{a} .
\end{equation}
In the configuration space, the corresponding Friedmann equation
reads
\begin{equation}\label{friedman}
- \frac{3}{8 \pi G} a \dot{a}^2 + \rho_0 a^{-3 w} = 0,
\end{equation}
and hence the scale factor runs as
\begin{equation}\label{scale-factor}
a(t) \propto (t_{br} \pm t)^{- \frac{2}{3( |w| - 1)}}
\end{equation}
where $t_{br}$ is a constant, and the $+$ and $-$ signs stand for
the quintessence and phantom regimes, respectively. The energy
density goes then as
\begin{equation}\label{energy density}
\rho \propto a^{3(|w|-1)}
\end{equation}
so, in the phantom case, $t_{br}$ turns out to be the time at the
so-called big rip singularity \cite{Caldwell}, where the scale
factor and the energy density blow up to infinity (see Fig.
\ref{univ}).

If one had to look at this phantom universe as evolving from an
initial coincidence time to infinity, then in order for the scale
factor, $a$, to be positive in (\ref{scale-factor}), not all
values of the parameter $w$ entering the equation of state for the
universe would be allowed. Actually, for a phantom regime only the
values given by \cite{Gonzalez-Diaz:2005sh}
\begin{equation}\label{discretization}
w= -\frac{1}{3}\left(1+\frac{2n+3}{n+1}\right), \;\; n=0, 1, 2,
... \infty       ,
\end{equation}
are allowed as the scale factor becomes ill-defined at $t>t_{br}$
otherwise. Moreover, for observers staying at times before the big
rip the singular character of this would necessarily imply that
the big rip should be cut off from the considered physical
manifold, so making completely unphysical the region beyond the
big rip singularity. Thus, besides introducing the condition
(\ref{discretization}) one must also take into account either that
stable wormholes and ringholes (whose existence is induced by the
violation of the dominant energy condition implied by the phantom
nature of the cosmic fluid) would crop up at both sides of the
singularity shortcutting the space-time \cite{Gonzalez03}, so
keeping the singularity outside the trajectories followed by
physical signalling and making therefore accessible to any
observers the contracting region beyond the big rip, no matter
whether it is cut off or not; or even by considering that quantum
effects could somehow smooth out the singular character of the big
rip. The condition (\ref{discretization}) could be thus
interpreted as a condition for the existence of a multiversal
scenario in which every value of $n$ would provide us with a
different universe with its own time evolution for the scale
factor. It is worth considering that this multiversal scenario can
still be a classical one, though its quantum counterpart presents
some interesting features that we will discuss later on.

\section{The quantum dark energy universe}

\subsection{The wave function}

Following the canonical quantization procedure, the dark energy
dominated universe would be quantum-mechanically characterized, in
the configuration space, by a wave function, $\Phi(a)$, which is
annihilated by the Hamiltonian density (\ref{Hamiltonian
Constraint}), in its operator form, satisfying in this way the
quantum version of the Hamiltonian constraint. But in doing that
we have to make a particular choice of the factor ordering between
the conjugate variables. There are many ways in which one can
depict this ambiguity, maybe a quite general one would be the
following. Classically, the conjugate variables commute so the rhs
and lhs in the following expression are equivalent
\begin{equation}\label{factor ordering}
\frac{p_a^2}{a} \equiv a^{-(r + s + 1)} p_a a^r p_a a^s ,
\end{equation}
but, of course, their quantum counterparts are not necessarily
equivalent as they depend on the commutation relations between the
conjugate variables. For instance, for the kinetic term of the
Hamiltonian (\ref{Hamiltonian Constraint}), we would obtain
\begin{equation}
\frac{\hat{p}_a^2}{\hat{a}} \rightarrow \frac{1}{\hat{a}}
\hat{p}_a^2 - \frac{(1- 2 \alpha) [\hat{a},\hat{p}_a]}{\hat{a}^2}
\hat{p}_a + \frac{\beta^2 [\hat{a},\hat{p}_a]^2}{\hat{a}^3},
\end{equation}
where we have used that
\begin{equation}
[ a^l , p_a ] = l [ a,p_a ] a^{l-1} ,
\end{equation}
and $\alpha$ and $\beta$ are related to the exponents in Eq.
(\ref{factor ordering}) through
\begin{eqnarray} \label{s}
s = - \alpha \pm \sqrt{\alpha^2 - \beta^2} , \\ \label{r} r = 1
\mp \sqrt{\alpha^2 - \beta^2} ,
\end{eqnarray}
so under canonical quantization, $p_a \rightarrow -i \hbar
\frac{\partial}{\partial a}$, $\alpha$ and $\beta$ represent the
factor ordering ambiguity in the corresponding Wheeler-DeWitt
equation \cite{DeWitt67}, which for the case being considered is
\begin{equation}\label{Wheeler-DeWitt dark energy}
\mathcal{N} \left( \frac{1}{a} \partial_{aa}^{2} +
\frac{1-2\alpha}{a^{2}}
\partial_{a} + \frac{\beta^{2}}{a^{3}} + \frac{\lambda_{0}^2}{\hbar^2} a^{-3w} \right) \Phi(a)
= 0 ,
\end{equation}
where $\lambda_0^2 = \frac{3}{2 \pi G} \rho_0 $, and $\mathcal{N}$
is the lapse function. Here, as well as in Eq. (\ref{Hamiltonian
Constraint}), we have assumed: (i) that dark energy dominates in
such a way that all matter in the universe is subdominant and can
therefore be disregarded, and (ii) that the quantum state is given
as a pure state. The first assumption stems from the feature that
the 70\% of the universal energy is made up of dark energy, even
though other fields should be considered in a more detailed
framework. The second assumption can be thought of as a plausible
one in spite of the state derived from it is not the most general.
The solutions of the Wheeler-DeWitt equation, Eq.
(\ref{Wheeler-DeWitt dark energy}), can be given in terms of
Bessel functions
\begin{equation} \label{funcion de ondas}
\Phi(a) = a^{\alpha} \, \mathcal{C}_{\nu} (\lambda a^{q}) ,
\end{equation}
where
\begin{eqnarray}\label{parametros funcion ondas}
q = \frac{3}{2} (1-w)  , \, \lambda = \frac{\lambda_0}{\hbar \, q}
, \ \nu^{2} q^2 = \alpha^2 - \beta^2 ,
\end{eqnarray}
$\mathcal{C}$ is the Bessel function of the first or second kind,
$\mathcal{J}_{\nu}$ and $\mathcal{Y}_{\nu}$, respectively, and $w$
satisfies the discretization (\ref{discretization}) for the case
of the phantom multiverse, and where we have kept the Planck
constant. This wave function would quantum-mechanically describe a
quintessence energy dominated universe, and actually the state of
the quantum universe for the phantom regime, were it not for the
possible existence of a noncausal multiply connected region around
the big rip singularity in the phantom energy dominated universe,
at least for the fraction of physical reality which always be
outside the connection. That wave function must satisfy, as usual,
given boundary conditions. For these boundary conditions we
choose: (i) the wave function ought to be regular everywhere, even
when the metric degenerates at time $t\rightarrow\infty$ in the
phantom universe, and (ii) it should vanish at the big rip
singularity when $a\rightarrow\infty$. These conditions are
satisfied by the wave function (\ref{funcion de ondas}) if we
enforce $\alpha$ to be
\begin{equation}\label{contorno1}
\alpha < \frac{q}{2} ,
\end{equation}
for the quintessence regime for $w > -1$ and for $w < -1$ just in
the region before the big rip, $t_{br}$. In the latter case, for
times after the singularity we must add the condition
\begin{equation}\label{contorno2} \alpha \pm q \, Re(\nu)
> 0 ,
\end{equation}
where the $+$ sign stands for the Bessel function of the first
kind, $\mathcal{J}_{\nu}$, and the $-$ sign does for the Bessel
function of the second kind, $\mathcal{Y}_{\nu}$. A linear
combination of these two solutions together with the boundary
conditions (\ref{contorno1}) and (\ref{contorno2}) would
represent, therefore, the quantum state of a dark energy dominated
universe. From this state we should be able to recover the
semi-classical universe in which we live. This may be accomplished
by taking the limit $\hbar \rightarrow 0$ in the expression for
the wave function. In particular, using the asymptotic expansion
of the Bessel functions \cite{Abramowitz72} for the wave function
(\ref{funcion de ondas}), we can obtain that in the semiclassical
approximation
\begin{equation}\label{semi-clasica}
\Phi \sim \sqrt{\frac{2}{\pi \lambda_0}} a^{(\alpha -
\frac{1}{2}q)} e^{\pm i (\lambda \, a^{q} - \frac{1}{2} \nu \pi -
\frac{1}{4} \pi)} .
\end{equation}
This wave function represents the state of the classical universe
in the sense that it is a quasi-oscillatory wave function whose
argument is essentially the classical action ($S_0 = \lambda
a^q$), so that the correlations between the classical variables
\cite{Halliwell87} are satisfied, i.e, $p_a = \frac{\partial
S_0}{\partial a}$, where $p_a$ is the classical momentum, is the
equation of motion, and
\begin{equation}
 \Delta = a^{\alpha - \frac{1}{2} q}
\end{equation}
in Eq. (\ref{semi-clasica}) is a prefactor smooth enough to
satisfy the Hartle criterion \cite{Hartle90}. In fact,
\begin{equation}
\mathcal{G}_{ijkl} \frac{\delta}{\delta h_{ij}}( |\Delta|^{2}
\frac{\delta S_{0}}{\delta h_{kl}} )=0 ,
\end{equation}
which in our case implies
\begin{equation}\label{hartcriterion dark energy}
\frac{1}{a} \; \frac{\partial}{\partial a} \left( a^{2 \alpha - q}
\frac{\partial S_0}{\partial a} \right)  \sim a^{2\alpha - 3}
\rightarrow 0 ,
\end{equation}
that is, the Hartle criterion is satisfied for quintessence models
($w>-1$) for the boundary condition (\ref{contorno1}). For the
phantom regime we should replace $2 \alpha < q$ for $2 \alpha < 3$
for the boundary condition, so that Eqs. (\ref{contorno1}) and
(\ref{hartcriterion dark energy}) are both satisfied. In both
cases, the semiclassical approximation (\ref{semi-clasica}) obeys
the Hamilton-Jacobi equation in the limit $\hbar \rightarrow 0$,
irrespective of the value of $q$ in the phantom multiverse that is
described by (\ref{discretization}), and for all choices of factor
ordering satisfying the boundary conditions. In particular, any
universe in this phantom multiverse would have a semi-classical
domain described by (\ref{semi-clasica}).

\subsection{The density matrix}

The most general quantum state would be given however in terms of
a mixed density matrix rather than a pure wave function
\cite{Hawking87}. One can also compute the density matrix for the
case being considered by taking
\begin{equation}\label{density matrix}
\rho (a',a) = \int_0^{\infty} dT \, K(a',T; a,0) ,
\end{equation}
where $K(a',T; a,0)$ is the Schr\"{o}dinger propagator and the
integration over time is introduced to account for the invariance
of the time separation between any two hypersurfaces.

In the case of our dark energy universe, we can take the gauge
$\mathcal{N} = a^3$ in Eq. (\ref{Wheeler-DeWitt dark energy}), so
we get a set of Hamiltonian eigenfunctions
\begin{equation}\label{eigenstates}
\hat{H}  \Phi_{k}(a)   =  \beta^2_{k}  \Phi_{k}(a)
\end{equation}
given by
\begin{equation}\label{hamiltonian eigenfunctions}
\Phi_k(a) = N_k \, a^{\alpha} \, \mathcal{J}_{k} (\lambda a^{q}) ,
\end{equation}
where $N_k$ is a normalization factor, and $q$ and $\lambda$ are
given by Eq. (\ref{parametros funcion ondas}), with eigenvalues
\begin{equation}\label{eigenvalues}
\beta^2_{k} = q^2 k^{2} - \epsilon^2_0 ,
\end{equation}
where $\epsilon_0^2 = \alpha^2 - \beta^2 \geq 0$ for the
parameters $r$ and $s$ in Eq. (\ref{factor ordering}) to be real
(see Eqs. (\ref{s}) and (\ref{r})).

However, this set of eigenfunctions are not orthogonal. For
instance, choosing for the scalar product
\begin{equation}\label{scalar product}
\langle f | g \rangle = \int_0^{\infty} da \, W(a) \, f(a) g(a),
\end{equation}
weighted by the function $W(a) = a^{-(2\alpha+1)}$, the set of
eigenfunctions would satisfy the following normalization relations
\begin{equation}\label{normalizationk}
\langle \Phi_k(a) | \Phi_k(a) \rangle = 1 , \forall k
> 0,
\end{equation}
in which we have used in Eq. (\ref{hamiltonian eigenfunctions})
$N_k = \sqrt{2 q k}$; In this way, for $ k \neq l$, we obtain
\begin{equation}\label{orthogonality}
\begin{array}{lc}
\langle \Phi_k(a) | \Phi_l(a) \rangle = 0   & , (k - l) \; \; \;
\textmd{even} \\ & \\
\langle \Phi_k(a) | \Phi_l(a) \rangle = \frac{4}{\pi}
\frac{(-1)^{\frac{1}{2}(k-l-1)} \sqrt{k \, l}}{k^2 - l^2} & , (k -
l)  \; \; \; \textmd{odd}.
\end{array}
\end{equation}
This set of equations can be considered as orthogonality relations
in the sense that they permit to split the whole Hilbert space
spanned by the Hamiltonian eigenfunctions into two Hilbert
subspaces, i.e., the subspaces spanned by the odd and even modes.
In that case, the Hamiltonian eigenfunctions form two orthogonal
basis, in the usual sense, for the subspaces.

But the zero mode is not normalizable. We may regularize it by
using some cut-off or minimum length, $l_p$, taking on the limit
$l_p \rightarrow 0$ at the end of the calculations. We have (see
the Appendix),
\begin{equation}
\langle \Phi_0 | \Phi_0 \rangle = \lim_{k,l \rightarrow 0} \langle
\Phi_k | \Phi_l \rangle \sim \frac{N_0^2}{q} \ln(\frac{2}{\lambda
l_p^q}) + \mathcal{O}(k \pm l),
\end{equation}
with which we could take the normalization relations
(\ref{normalizationk}) for all $k \geq 0$, with a suitable
normalization factor, $N_0$.

With that regularization, we can make use of the following set of
 functions
\begin{equation}\label{orthonormal basis}
\Psi_n(a) = \sqrt{q \lambda} \, a^{\alpha + \frac{1}{2} q} \, e^{-
\frac{\lambda a^q}{2}} L_n(\lambda a^q) ,
\end{equation}
which are constructed in terms of the Laguerre polynomials,
$L_n(x)$,
\begin{equation}
L_n(x) = \sum_{m=0}^n \left( \begin{array}{c} n
\\ m\end{array}\right) \frac{(-x)^m}{m!} .
\end{equation}
We have in this way obtained an orthonormal set under the scalar
product (\ref{scalar product}), which can be used as the basis for
the square integrable functions upon which the Hamiltonian would
act, so
\begin{equation}
\sum_n |\Psi_n \rangle \langle \Psi_n | = Id .
\end{equation}
Using this set, we have
\begin{equation}\label{identity partition}
Id = \sum_n | \Psi_n\rangle \langle \Psi_n| = \sum_{k l}
\mathcal{D}_{k l} |\Phi_k\rangle \langle \Phi_k|
\end{equation}
where
\begin{equation}\label{coeficientesD}
D_{k l} = \sum_n C_{n k} C^*_{n l} ,
\end{equation}
and the $C_{i j}$'s are the coefficients for the change of basis,
i.e.,
\begin{equation}\label{expansionPsi}
\Psi_n (a) = \sum_m C_{n m} \Phi_m(a) .
\end{equation}
Hence, the propagator can be written as
\begin{eqnarray} \nonumber
K(a',T;a,0) & = & \sum_{k l} D_{k l} \langle a' |
e^{\frac{i}{\hbar} T H} | \Phi_k \rangle \langle \Phi_l | a
\rangle \\ \label{propagator} & = & \sum_{k l} D_{k l}
e^{\frac{i}{\hbar} T \beta^2_k} \Phi_k(a') \Phi_l^*(a) ,
\end{eqnarray}
and the density matrix computed to be
\begin{equation}\label{density matrix dark energy}
\rho(a',a) = \sum_{k l} D_{k l} \frac{\Phi_k(a') \Phi_l(a)}{q^2
k^{2} - \epsilon^2_0} ,
\end{equation}
where, in order to make the integral well-defined, we have Wick
rotated time counterclockwise. Wick rotating in the opposite
direction would have implied inserting the identity before the
evolution operator and setting a minus sign in the exponent.

In the case of the dark energy universe, with the Hamiltonian
eigenfunctions (\ref{hamiltonian eigenfunctions}), the
coefficients in Eqs. (\ref{coeficientesD}) and
(\ref{expansionPsi}) can be computed. For, we can take advantage
of the properties of the first kind Bessel functions, which form
up an overcomplete set in the sense that any arbitrary function
can be decomposed in terms of Bessel functions through a Neumann's
expansion \cite{Watson}
\begin{equation}\label{Neumann expansion}
z^{\nu}f_k(z) = \sum_{n=0}^{\infty} c_{kn} \mathcal{J}_{n+\nu}(z)
,
\end{equation}
with the coefficients being given by
\begin{equation}\label{coefficients}
c_{kn} = \frac{1}{2 \pi i} \int_{|t|<R} dt \, f_k(t) A_{n,\nu}(t)
,
\end{equation}
where $R$ is the distance from $t=0$ to the closest pole of
$f_k(t)$, and the $A_{n,\nu}$, the Gegenbauer's polynomials, are
defined by
\begin{equation}
A_{n,\nu}(t) = \frac{2^{n + \nu} (n + \nu)}{t^{n+1}}
\sum_{m=0}^{\leq \frac{1}{2}n} \frac{\Gamma(n + \nu -m)}{m!}
\left( \frac{t}{2}\right)^{2m} .
\end{equation}
Thus, we can rearrange Eq. (\ref{Neumann expansion}) so that any
arbitrary function could be written as
\begin{equation}
g_k(a) = a^{\alpha} \left( \lambda a^q \right)^{\nu} f_k(\lambda
a^q) = \sum_{n=0}^{\infty} c_{kn}
 a^{\alpha} \mathcal{J}_{n + \nu}(\lambda a^q) .
\end{equation}
In particular, with $\nu = \frac{1}{2}$, the orthonormal set of
functions (\ref{orthonormal basis}) can be decomposed as
\begin{equation}\label{expansion}
\Psi_k(a) = \sum_{n=0}^{\infty} C_{k n} \Phi_{n+\frac{1}{2}}(a)
\end{equation}
in which the coefficients, unless for a normalization constant,
are given by Eq. (\ref{coefficients}), with
\begin{equation}
f_k(t) = \sqrt{q} \; e^{-\frac{t}{2}} \sum_{l=0}^{k} \left(
\begin{array}{c} k \\ l \end{array}\right) \frac{(-1)^l}{l!} t^l ,
\end{equation}
that is,
\begin{widetext}
\begin{eqnarray} \nonumber
C_{k n} & =& \frac{ (n+\frac{1}{2}) \, 2^{n+\frac{1}{2}}
}{N_{n+\frac{1}{2}}} \frac{1}{2 \pi i} \sqrt{q} \, \sum_{l=0}^k
\sum_{m=0}^{\leq \frac{1}{2}n} \left(
\begin{array}{c}k\\l \end{array} \right) \frac{(-1)^l \; \Gamma(n+\frac{1}{2}-m)}{2^{2m} \; l! \; m!} \, \int_{|t|<R} \frac{e^{-\frac{t}{2}}}{t^{n-2m-l+1}} dt \, \\
\label{coeficientesC} & = & \frac{ (-1)^{n} \; (n+\frac{1}{2}) \,
2^{n+\frac{1}{2}} }{N_{n+\frac{1}{2}}} \sqrt{q} \, \sum_{m \geq
\frac{n - k}{2}}^{\leq \frac{1}{2}n} \left(
\begin{array}{c}k\\n - 2m \end{array} \right) \frac{\Gamma(n+\frac{1}{2}-m)}{2^{2m} \; (n - 2m)! \; m!}
,
\end{eqnarray}
\end{widetext}
where $m$ is a non-negative integer. The density matrix
(\ref{density matrix dark energy}) can then be written as
\begin{equation}\label{density matrix dark energy2}
\rho(a',a) = \sum_{k l} D_{k l} \frac{\Phi_{k+\frac{1}{2}}(a')
\Phi_{l+\frac{1}{2}}(a)}{q^2 (k+\frac{1}{2})^{2} - \epsilon^2_0} .
\end{equation}

A density matrix for a physical system is supposed to be definite
positive, and Eq. (\ref{density matrix dark energy2}) is however
not necessary so, even for positive values of the coefficients
$D_{kl}$. Parameters $\alpha$ and $\beta$ have nevertheless no
clear physical meaning since a semiclassical state should be
independent of the particular choice of $\alpha$ and $\beta$. We
then could still take for the physical state of the universe the
reduced density matrix resulting from integrating out $\alpha$ and
$\beta$, i.e.,
\begin{equation}
\rho_r(a',a) = \int d\alpha \int d\beta \, \rho(a',a;\alpha,\beta)
,
\end{equation}
This integral turns out to be divergent so that one could not
obtain a meaningful density matrix for the state of the universe.
One way to avoid this problem could be taking some particular
values for the factor ordering as a boundary condition. For the
particular choice $\alpha = \beta = 0$ we would in fact have
\begin{equation}\label{density matrix particular choice}
\rho (a', a) = \sum_{n m} D_{nm} \frac{\Phi_{n+\frac{1}{2}}(a')
\Phi_{m+\frac{1}{2}}(a)}{ q^2 (m+\frac{1}{2})^2},
\end{equation}
where the coefficients are given by Eq. (\ref{coeficientesD}).
However, even though this density matrix does not show the usual
divergences due to vanishing values of the denominator, the
coefficients $D_{nm}$ in it are still divergent.

On the other hand, our particular choice leading to the density
matrix (\ref{density matrix particular choice}) corresponds to a
rather arbitrary choice of the factor ordering. Other choices
could also imply a definite positive density matrix. It would
follow that an alternate philosophy could be constructing a
propagator in terms of pairs of levels instead of the single
levels which correspond to the Hamiltonian eigenvalues, that is
\begin{equation}
\langle a' | e^{i T H} | a \rangle = \sum_{nm} D_{nm} \, \langle
a' | e^{i \frac{T H}{2}} | \Phi_{n+\frac{1}{2}} \rangle \langle
\Phi_{m+\frac{1}{2}} | e^{i \frac{T H}{2}} | a \rangle ,
\end{equation}
In that case, we have
\begin{equation}\label{density matrix without factor ord.}
\rho (a', a) = \sum_{n m} D_{nm} \frac{\Phi_{n+\frac{1}{2}}(a')
\Phi_{m+\frac{1}{2}}(a)}{ q^2 (m^2 - n^2 + m - n )} ,
\end{equation}
i.e., although the factor ordering ambiguity is no longer present
in the denominator, the diagonal elements become now divergent.

The kinds of divergences and unphysical states that we have just
uncovered were already pointed out by Hawking and Page for density
matrix in quantum cosmology \cite{Hawking87} \cite{Page}.
Actually, these difficulties can be seen to arise because the
system can reach a state with zero value for its Hamiltonian
eigenvalue, and would be expected to be solved in the framework of
a proper quantum theory of gravity, in which a minimum energy, the
Planck mass, $m_p$, should exist. In such a case, a nonzero
minimum value of the Hamiltonian eigenvalue would be expected that
rendered the density matrix definite positive and always
convergent.

\subsection{Entangled states in the multiverse}

Let us, now, be concerned with a phantom universe and its big rip
singularity, then we could consider the case in which no wormholes
are connecting the regions before and after the big rip; i.e. when
the wormholes that branch off in the neighborhood of that
singularity simply connect two asymptotic regions on the same side
of the singularity. If thereby such wormholes are disregarded and
the hypersurface at the singularity is cut out so that the whole
space-time is divided into two separate parts, then there will be
two independent wave functions which should be associated with
different realizations of the boundary conditions and distinct
time intervals. The first of these intervals runs from the
coincidence time until the time at the big rip, and the second one
goes from the latter time until infinity.  The general boundary
conditions that the quantum state be regular everywhere and
exactly vanishes at the big rip singularity amount to a wave
function which should be generally expressed in terms of a
different linear combinations of first and second kind Bessel's
functions, $\mathcal{J}$ and $\mathcal{Y}$, on each interval.
These two wave functions for both sectors can be regarded to play
the role of some bases for the quantum state of a specific
$n$-phantom universe. So, in general we can describe this state as
\begin{equation}
\Psi_n=c_I^n\Psi_I^n +c_{II}^n\Psi_{II}^n ,
\end{equation}
with
\begin{equation}
\left(c_I^n\right)^2 +\left(c_{II}^n\right)^2 = 1 .
\end{equation}

Let us interpret for a moment the integer number $n$ defined in
Eq. (\ref{discretization}) as a quantum number labeling the
different universes in our multiverse, and then consider the
quantum states for two universes with different values of the
quantum number $n$. In that case, $\psi_I^n$ and $\psi_{II}^n$,
which quantum-mechanically describe the regions before and after
the singularity for a single universe labeled $n$, are both
strongly peaked at time-like separated regions and can therefore
be correlated to the similar regions of the other universe, say
$m$, as all of such states satisfy the so called Hartle criterion
(\ref{hartcriterion dark energy}). The result for the two
universes could be a common state
\begin{equation}\label{entangled state in multiverse}
\Phi=c_I^n c_I^m\Psi_I^n\Psi_I^m + c_{II}^n
c_{II}^m\Psi_{II}^n\Psi_{II}^m .
\end{equation}
Since the singularity has been cut off, mixed states $\psi_I^n
\psi_{II}^m$ and $\psi_{II}^n \psi_{I}^m$ are no longer possible
if there are correlations between $\psi_I^n \psi_{I}^m$ and
$\psi_{II}^n \psi_{II}^m$ and, therefore, state (\ref{entangled
state in multiverse}) should be an entangled state.

Eq. (\ref{entangled state in multiverse}) can straightforwardly be
generalized to the infinite possible number of universes and would
imply that knowing the state of our universe we would
automatically know the state of the other universes belonging to
the same multiverse scenario. This is a cosmic translation from
what is usually dubbed quantum non-locality in quantum information
theory. Since there is no space-time between any two universes in
this multiverse, the term locality (or non-locality) used to
characterize correlations between two particles in a common
space-time, is no longer suitable to physically characterize the
above correlations between universes, provided that locality
refers to just space-like or time-like location in a common
space-time. It would instead refer to correlations between the
quantum states of different universes, which would become
entangled as a result.

\section{Generalized phantom universe}

However, the existence of a non-chronal region around the big rip
makes it impossible to have any quantum state for the phantom
universe. Actually, in order for having a proper quantum theory of
one of the universes of the multiverse we have to make use of a
generalized quantum theory \cite{Hartle90}. Technically, the
physical system we ought to deal with consists of a space-time
manifold containing an intermediate bounded non-foliating region
on the neighborhood of the big rip singularity, filled with closed
time-like curves (CTCs), which is chronologically placed between
two regions that are both foliable by a family of nonintersecting
spacelike surfaces $\Sigma_p$ (See Fig. 1). We then introduce the
generalized decoherence function of Hamiltonian mechanics which
reads
\begin{widetext}
\begin{equation}\label{generalized deco}
D(\alpha ',\alpha)= N{\rm Tr}\left[P^p_{\alpha
'_{p}}(\Sigma_p)...P^{k+1}_{\alpha '_{k+1}}(\Sigma_{k+1}) X
P^{k}_{\alpha'_{k}}(\Sigma_{k})...P^{1}_{\alpha'_{1}}(\Sigma_{1})\rho
P^{1}_{\alpha_{1}}(\Sigma_{1})...P^{k}_{\alpha_{k}}(\Sigma_{k})X^{\dagger}
P^{k+1}_{\alpha_{k+1}}(\Sigma_{k+1})...P^p_{\alpha_{p}}(\Sigma_p)\right]
,
\end{equation}
\end{widetext}
where the $P_{\alpha}$'s are projection operators which forms up a
set, $\{P_{\alpha_{p}}\}$, that corresponds to the exhaustive and
exclusive set of alternatives defined on a given non-intersecting
spacelike surface $\Sigma$, and the $\alpha_{j}$'s are particular
sequences of coarse-grained alternatives
$\{\alpha\}=\alpha_1,...,\alpha_j$ that describe particular
histories. The exhaustive set of histories consists then all
possible sequences $\{\alpha\}$. The function $\rho$ denotes the
density matrix encompassing the boundary condition of the system
on an initial nonintersecting spacelike surface $\Sigma_0$. It
will be given either by the factorizable probability
$\rho=W=\Phi(a)\Phi(a')$ or by the expressions for the density
matrix described in the previous section, if the probability
function $W$ is not factorizable, i.e., for a mixed state. $X$ is
a generalized evolution matrix that can be defined in terms of a
nonunitary matrix $X_s$ which replaces the usual unitary evolution
matrix $U$ of the decoherence function for fully foliable
manifolds. It can be given by
\begin{equation}
X=U\left(\Sigma_f, \Sigma_{\infty})^{-1}X_s U(\Sigma_i,
\Sigma_0\right)  ,
\end{equation}
in which $\Sigma_0$ and $\Sigma_{\infty}$ are surfaces at the
furthest possible past and future, the latter being assumed to be
at the event horizon (i.e. in the present case at infinity), and
$\Sigma_i$ and $\Sigma_f$ are the latest and earliest
nonintersecting spacelike surfaces, after and before the
nonfoliable region, respectively. Finally, the normalizing factor
$N$ is given by
\begin{equation}
N=\left[{\rm Tr}(X\rho X^{\dagger})\right]^{-1} .
\end{equation}
It can be then shown that the generalized decoherence function
(\ref{generalized deco}) is normalizable and hermitian, has
positive diagonal elements, and satisfies the superposition
principle, provided that \cite{Hartle94}
\[D(\bar{\alpha}',\bar{\alpha})= \sum_{\alpha
'\in\bar{\alpha}'}\sum_{\alpha\in\bar{\alpha}}D(\alpha ',\alpha)
\]
for all coarse grainings $\{\bar{\alpha}\}$ of $\{\alpha\}$. Thus,
this function satisfies all consistency tests.

In the sum-over-histories formulation of gravitational systems the
usual probability function $W$ is replaced for a probability
function for a given set of alternatives $\alpha$, $p(\alpha)$,
which can be obtained from the decoherence function by using the
relation $D(\alpha ',\alpha)\approx\delta_{\alpha
'\alpha}p(\alpha)$. We shall take these probabilities $p(\alpha)$
as the physical quantities that replace quantum states in our
non-causal system. So, we are here particularly interested in
calculating from Eq. (\ref{generalized deco}) the probability
$p(\alpha,\Sigma;t_0,a)$ of a set of alternatives for particular
values of $t_0$, $a$ and $P_{\alpha}\left[B(t_0,a)\right]$, that
distinguish only the scale factor values on the whole pieces of
surfaces , $B(t_0,a)$, which should be spacelike separated from
the nonfoliating region. The most general expression for the
probability $p(t_0,a)$ (in which we have specialized at the
particular slicings $t=t_0$ and $a=a$) that can be obtained from
the decoherence function (\ref{generalized deco}) is
\begin{widetext}
\begin{eqnarray}\label{probabilitiesParcial}
&&p\left(\alpha,\Sigma ';t_0,a\right)\equiv p\left(\alpha,\Sigma
'';t_0,a\right)\nonumber\\ &&=N{\rm Tr}\left\{X
P_{\alpha}\left[B(t_0,a)\right]\rho
P_{\alpha}\left[B(t_0,a)\right]X^{\dagger}\right\}\equiv N{\rm
Tr}\left\{P_{\alpha}\left[B(t_0,a)\right]X\rho
X^{\dagger}P_{\alpha}\left[B(t_0,a)\right]\right\} .
\end{eqnarray}
\end{widetext}
Finally, the quantity
\begin{equation}\label{probabilitiesAlternatives}
p(\alpha,\Sigma ')\equiv p(\alpha,\Sigma '')= \int_{0}^{\infty}
dt_0 \int_{0}^{\infty} da p\left(\alpha,\Sigma^{(\upsilon)};
t_0,a\right) ,
\end{equation}
where the superscript $(\upsilon)$ denotes either $'$ or $''$, is
the most general probability for the set of all alternatives that
are able to distinguish the scale factor values on spacelike
sections that are spacelike separated from the identified
nonfoliating region. It will be here regarded to be the quantity
that replaces the quantum state for a parallel phantom universe in
Hamiltonian quantum cosmology.

Because of the non-unitarity of the evolution in the neighborhood
of the big rip singularity, described in Eq.
(\ref{probabilitiesParcial}) by the non-unitary operator $X$, the
probabilities given by Eq. (\ref{probabilitiesAlternatives}) for
alternative histories completely defined on a current local piece
located around e.g. our galaxy on a given hypersurface, would
depend on the state of the acronal region \cite{Hartle94}, in the
same way as probabilities for retrodiction histories
\cite{Hartle90} depend on the current and initial states of the
universe. In fact, by the cyclic property of the trace, we can
rewrite Eq. (\ref{probabilitiesParcial}) as
\begin{equation}\label{causality}
p\left(\alpha,\Sigma ';t_0,a\right) = N{\rm Tr}\left\{ \rho_f
P_{\alpha}\left[B(t_0,a)\right]\rho
P_{\alpha}\left[B(t_0,a)\right]\right\} .
\end{equation}
where the density matrix, $\rho_f$, is given by $\rho_f = X
X^{\dagger}$. It would mean that, experiments in local
laboratories, say the solar system or the Galaxy, might give
different results depending on whether this non-chronal region in
our future exists or not. So, at least from a theoretical point of
view, it could glimpse the idea of measuring global properties
from \emph{ local} experiments.

\section{Conclusions}

A multiverse scenario can arise in the realm of a phantom energy
dominated universe when we consider the complete range of the time
interval, smoothing somehow the big rip singularity. That
multiverse scenario comes up from the constraint that we need to
impose onto the equation of state parameter, in order to obtain
well-defined values for the scale factor at times after the
singularity.

Analytic solutions can be obtained for the quantum state of a dark
energy dominated universe. If the state of the universe is given
by a wave function, i.e., if it is a pure state, its quantum
representation can be expressed as a linear combination of Bessel
functions and, imposing the appropriate boundary conditions, we
can recover the semiclassical approximation in the usual way.
Nevertheless, the most general quantum state for the universe
should be given by a density matrix, which not with understanding
suffers from the usual divergence shortcomings. In order to
compute an explicit expression for the density matrix in the case
being considered, we use a particular gauge in which the
Hamiltonian eigenfunctions can be found, and that can be employed
as a basis for the space of functions that the Hamiltonian acts
upon, although this basis is not orthogonal. Then, we found an
orthonormal set in terms of which the Hamiltonian eigenfunctions
can be expressed and, thus, several expressions for the density
matrix are given.

We show that cosmic entangled states between universes can take
place in the realm of the phantom multiverse. We also give a
quantum description of a phantom universe when the singularity is
replaced for a bounded non-chronal region. In such a case, the
generalized quantum theory is applied and consistent expressions
for the probabilities of alternative histories are given.

Although we have succeeded in obtaining a function that replaces
the conventional notion of quantum state in a cosmological
spacetime endowed with a bounded multiply connected region, the
model considered in this paper is not realistic enough for at
least the following reason. We have not specifically introduced
any matter fields in the model, so that this should at best be
considered as an asymptotic idealization.

Now some kinds of quantum communication channels could be
conceived which related the different universes that belong to the
multiverse. Whether these communications could be implemented
physically between advanced civilizations existing in such
universes is a matter that requires further consideration.

\acknowledgements The authors thank M. P. Mart\'{\i}n-Moruno and
A. Rozas-Fern\'{a}ndez for support. This paper was supported
partly by CAICYT under Research Project Nº FIS2005-01181.

\appendix

\section{Orthogonality properties of Bessel functions}

The usual formula for the integrals of Bessel functions can be
obtained from the standard bibliography, to be \cite{Abramowitz72}
\begin{widetext}
\begin{equation}
F(z) \equiv \int^z \frac{1}{t} J_{\mu}(k t) J_{\nu}(k t) \, dt = -
\frac{1}{\mu^2 - \nu^2} \{ k z \left[ J_{\mu+1}(k t) J_{\nu}(k t)
-  J_{\mu}(k t) J_{\nu+1}(k t) \right] - (\mu - \nu) J_{\mu}(k t)
J_{\nu}(k t) \}
\end{equation}
\end{widetext}
Then, a definite integral over t summing from $0$ to $\infty$ can
be thought of as the substraction of the two following limits,
\begin{equation}\label{integral}
\int_0^{\infty} dt \, \frac{1}{t}  J_{\mu}(k t) J_{\nu}(k t) =
\lim_{z \rightarrow \infty} F(z) - \lim_{z \rightarrow 0} F(z) .
\end{equation}
Let us compute $\lim_{z \rightarrow 0} F(z)$ first. In this case,
taking the asymptotic limits for the Bessel's functions, we have
\begin{widetext}
\begin{eqnarray}\nonumber
\lim_{z \rightarrow 0} F(z) & \approx & - \frac{1}{\mu^2 - \nu^2}
\{ k z \left[ \left(\frac{k z}{2} \right)^{\mu + \nu + 1} \left(
\frac{1}{\Gamma (\mu + 2) \Gamma(\nu + 1)} - \frac{1}{\Gamma (\mu
+ 1) \Gamma(\nu + 2)} \right) \right] \\ \nonumber & & - (\mu -
\nu) \left( \frac{k z}{2} \right)^{\mu + \nu} \frac{1}{\Gamma (\mu
+ 1) \Gamma(\nu + 1)} \} \\ \nonumber & = & \left( \frac{k z}{2}
\right)^{\mu + \nu} \frac{1}{(\mu - \nu) (\mu + \nu)} \left[ 2
\left( \frac{k z}{2} \right)^2 \frac{(\mu - \nu)}{\Gamma (\mu + 2)
\Gamma(\nu + 2)} + (\mu - \nu) \frac{1}{\Gamma (\mu + 1)
\Gamma(\nu + 1)} \right] \\ \label{limite cero} & \approx &
\frac{1}{\Gamma (\mu + 1) \Gamma(\nu + 1)} \, \frac{\left( \frac{k
z}{2} \right)^{\mu + \nu}}{\mu + \nu } \rightarrow 0 \; (\mu + \nu
> 0 ).
\end{eqnarray}
\end{widetext}
We can check that Eq. (\ref{limite cero}) vanishes in the limit
for any $(\mu + \nu > 0)$, but it should be taken into account if
a cut off at the Planck length is introduced to regularize the
zero mode Bessel function. For the upper limit we obtain
\begin{widetext}
\begin{eqnarray}\nonumber
\lim_{z \rightarrow \infty} F(z) & \approx & \frac{-1}{\mu^2 -
\nu^2} \frac{2 k}{\pi} [ \cos(z - \frac{1}{2} (\mu + 1) \pi -
\frac{\pi}{4}) \cos(z - \frac{1}{2} \nu \pi - \frac{\pi}{4}) \\ &
& \nonumber - \cos(z - \frac{1}{2} \mu \pi - \frac{\pi}{4}) \cos(z
- \frac{1}{2} (\nu+1) \pi - \frac{\pi}{4}) ] +
\mathcal{O}(\frac{1}{z}) \\ \label{limite infinito} & = &
\frac{k}{\mu + \nu} \, \frac{\sin(\frac{\pi}{2}(\nu
-\mu))}{\frac{\pi (\nu -\mu)}{2}}
\end{eqnarray}
\end{widetext}
For $\mu + \nu >0$, it gives:

1) if $\mu = \nu$
\begin{equation}
\lim_{z \rightarrow \infty} F(z) = \frac{k}{\mu + \nu}
\end{equation}
2) if $\mu \neq \nu$ and $\nu - \mu = 2n$ (even)
\begin{equation}
\lim_{z \rightarrow \infty} F(z) = 0
\end{equation}
3) if $\mu \neq \nu$ and $\nu - \mu = 2n+1$ (odd)
\begin{equation}
\lim_{z \rightarrow \infty} F(z) = \frac{2 k}{\pi} \,
\frac{(-1)^{\frac{1}{2}(\nu-\mu-1)}}{(\nu-\mu)(\nu+\mu)}
\end{equation}
Essentially, these are the values of the integral Eq.
(\ref{integral}) since the limit to zero vanishes. It is left the
case for the zero mode. We can regularize it by taking some
cut-off or minimum length, and evaluate the limit
\begin{equation}\label{a8}
\lim_{l_p \rightarrow 0}\lim_{\mu+\nu \rightarrow 0}
\int_{l_p}^{\infty} dt \, \frac{1}{t} J_{\mu}(k t) J_{\nu}(k t) ,
\end{equation}
expanding in $\mu$ and $\nu$ the limits Eq. (\ref{limite cero})
and Eq. (\ref{limite infinito}), i.e.
\begin{equation} \frac{1}{\mu + \nu} \,
\frac{\sin(\frac{\pi}{2}(\nu -\mu))}{\frac{\pi (\nu -\mu)}{2}}
\approx \frac{1}{\mu + \nu} (1 + \mathcal{O}((\mu -\nu)^2))
\end{equation}
and
\begin{equation} \frac{\left( \frac{l_p}{2} \right)^{\mu +
\nu}}{\mu + \nu } \approx \frac{1}{\mu + \nu} (1+ (\mu + \nu)
\ln(\frac{l_p}{2}) + \mathcal{O}((\mu + \nu)^2) )
\end{equation}
so, the limit in Eq. (\ref{a8}) gives
\begin{equation}
\lim_{\mu+\nu \rightarrow 0} \lim_{l_p \rightarrow 0}
\int_{l_p}^{\infty} dt \, \frac{1}{t} J_{\mu}(k t) J_{\nu}(k t) =
\lim_{l_p \rightarrow 0} \ln(\frac{2}{l_p}) .
\end{equation}

\end{document}